\documentstyle[12pt]{article}
\advance \topmargin by -\headheight
\advance \topmargin by -\headsep

\evensidemargin \oddsidemargin
\begin{document}

\topmargin -.6in

\newcommand{\sect}[1]{\setcounter{equation}{0}\section{#1}}
\renewcommand{\theequation}{\thesection.\arabic{equation}}
\relax

%
\def\rf#1{(\ref{eq:#1})}
\def\lab#1{\label{eq:#1}}
\def\nonu{\nonumber}
\def\br{\begin{eqnarray}}
\def\er{\end{eqnarray}}
\def\be{\begin{equation}}
\def\ee{\end{equation}}
\def\eq{\!\!\!\! &=& \!\!\!\! }
\def\foot#1{\footnotemark\footnotetext{#1}}
\def\lb{\lbrack}
\def\rb{\rbrack}
\def\llangle{\left\langle}
\def\rrangle{\right\rangle}
\def\blangle{\Bigl\langle}
\def\brangle{\Bigr\rangle}
\def\llb{\left\lbrack}
\def\rrb{\right\rbrack}
\def\lcurl{\left\{}
\def\rcurl{\right\}}
\def\({\left(}
\def\){\right)}
\newcommand{\nit}{\noindent}
\newcommand{\ct}[1]{\cite{#1}}
\newcommand{\bi}[1]{\bibitem{#1}}
\def\lskip{\vskip\baselineskip\vskip-\parskip\noindent}
\relax
\def\mskp{\par\vskip 0.3cm \par\noindent}
\def\sskp{\par\vskip 0.15cm \par\noindent}
\def\tr{\mathop{\rm tr}}
\def\Tr{\mathop{\rm Tr}}
\def\v{\vert}
\def\bv{\bigm\vert}
\def\Bgv{\;\Bigg\vert}
\def\bgv{\bigg\vert}
\newcommand\partder[2]{{{\partial {#1}}\over{\partial {#2}}}}
\newcommand\funcder[2]{{{\delta {#1}}\over{\delta {#2}}}}
\newcommand\Bil[2]{\Bigl\langle {#1} \Bigg\vert {#2} \Bigr\rangle}  
\newcommand\bil[2]{\left\langle {#1} \bigg\vert {#2} \right\rangle} 
\newcommand\me[2]{\left\langle {#1}\right|\left. {#2} \right\rangle} 
\newcommand\sbr[2]{\left\lbrack\,{#1}\, ,\,{#2}\,\right\rbrack}
\newcommand\pbr[2]{\{\,{#1}\, ,\,{#2}\,\}}
\newcommand\pbbr[2]{\lcurl\,{#1}\, ,\,{#2}\,\rcurl}
%
\def\a{\alpha}
\def\b{\beta}
\def\d{\delta}
\def\D{\Delta}
\def\eps{\epsilon}
\def\vareps{\varepsilon}
\def\g{\gamma}
\def\G{\Gamma}
\def\grad{\nabla}
\def\h{{1\over 2}}
\def\l{\lambda}
\def\L{\Lambda}
\def\m{\mu}
\def\n{\nu}
\def\o{\over}
\def\om{\omega}
\def\O{\Omega}
\def\p{\phi}
\def\P{\Phi}
\def\vp{\varphi}
\def\pa{\partial}
\def\pr{\prime}
\def\ra{\rightarrow}
\def\s{\sigma}
\def\S{\Sigma}
\def\t{\tau}
\def\th{\theta}
\def\Th{\Theta}
\def\ti{\tilde}
\def\wti{\widetilde}
\def\bp{{\bar \p}}
\newcommand\sumi[1]{\sum_{#1}^{\infty}}   
\newcommand\fourmat[4]{\left(\begin{array}{cc}  
{#1} & {#2} \\ {#3} & {#4} \end{array} \right)}
\newcommand\twocol[2]{\left(\begin{array}{c}  
{#1} \\ {#2} \end{array} \right)}

%
\def\lie{{\cal G}}
\def\dlie{{\cal G}^{\ast}}
\def\f#1#2#3 {f^{#1#2}_{#3}}
\def\winf{{\sf w_\infty}}
\def\win1{{\sf w_{1+\infty}}}
\def\hwinf{{\sf {\hat w}_{\infty}}}
\def\Winf{{\sf W_\infty}}
\def\Win1{{\sf W_{1+\infty}}}
\def\hWinf{{\sf {\hat W}_{\infty}}}
\def\cB{{\cal B}}
\def\cH{{\cal H}}
\def\cL{{\cal L}}
\def\cA{{\cal A}}
\def\cK{{\cal K}}
\def\cM{{\cal M}}
\def\cR{{\cal R}}
\def\cP{{\cal P}}
%
\def\rlx{\relax\leavevmode}
\def\inbar{\vrule height1.5ex width.4pt depth0pt}
\def\IZ{\rlx\hbox{\sf Z\kern-.4em Z}}
\def\IR{\rlx\hbox{\rm I\kern-.18em R}}
\def\IC{\rlx\hbox{\,$\inbar\kern-.3em{\rm C}$}}
\def\one{\hbox{{1}\kern-.25em\hbox{l}}}
%
%
\newcommand\PRL[3]{{\sl Phys. Rev. Lett.} {\bf#1} (#2) #3}
\newcommand\NPB[3]{{\sl Nucl. Phys.} {\bf B#1} (#2) #3}
\newcommand\NPBFS[4]{{\sl Nucl. Phys.} {\bf B#2} [FS#1] (#3) #4}
\newcommand\CMP[3]{{\sl Commun. Math. Phys.} {\bf #1} (#2) #3}
\newcommand\PRD[3]{{\sl Phys. Rev.} {\bf D#1} (#2) #3}
\newcommand\PLA[3]{{\sl Phys. Lett.} {\bf #1A} (#2) #3}
\newcommand\PLB[3]{{\sl Phys. Lett.} {\bf #1B} (#2) #3}
\newcommand\JMP[3]{{\sl J. Math. Phys.} {\bf #1} (#2) #3}
\newcommand\PTP[3]{{\sl Prog. Theor. Phys.} {\bf #1} (#2) #3}
\newcommand\SPTP[3]{{\sl Suppl. Prog. Theor. Phys.} {\bf #1} (#2) #3}
\newcommand\AoP[3]{{\sl Ann. of Phys.} {\bf #1} (#2) #3}
\newcommand\PNAS[3]{{\sl Proc. Natl. Acad. Sci. USA} {\bf #1} (#2) #3}
\newcommand\RMP[3]{{\sl Rev. Mod. Phys.} {\bf #1} (#2) #3}
\newcommand\PR[3]{{\sl Phys. Reports} {\bf #1} (#2) #3}
\newcommand\AoM[3]{{\sl Ann. of Math.} {\bf #1} (#2) #3}
\newcommand\UMN[3]{{\sl Usp. Mat. Nauk} {\bf #1} (#2) #3}
\newcommand\FAP[3]{{\sl Funkt. Anal. Prilozheniya} {\bf #1} (#2) #3}
\newcommand\FAaIA[3]{{\sl Functional Analysis and Its Application} {\bf #1}
(#2) #3}
\newcommand\BAMS[3]{{\sl Bull. Am. Math. Soc.} {\bf #1} (#2) #3}
\newcommand\TAMS[3]{{\sl Trans. Am. Math. Soc.} {\bf #1} (#2) #3}
\newcommand\InvM[3]{{\sl Invent. Math.} {\bf #1} (#2) #3}
\newcommand\LMP[3]{{\sl Letters in Math. Phys.} {\bf #1} (#2) #3}
\newcommand\IJMPA[3]{{\sl Int. J. Mod. Phys.} {\bf A#1} (#2) #3}
\newcommand\AdM[3]{{\sl Advances in Math.} {\bf #1} (#2) #3}
\newcommand\RMaP[3]{{\sl Reports on Math. Phys.} {\bf #1} (#2) #3}
\newcommand\IJM[3]{{\sl Ill. J. Math.} {\bf #1} (#2) #3}
\newcommand\APP[3]{{\sl Acta Phys. Polon.} {\bf #1} (#2) #3}
\newcommand\TMP[3]{{\sl Theor. Mat. Phys.} {\bf #1} (#2) #3}
\newcommand\JPA[3]{{\sl J. Physics} {\bf A#1} (#2) #3}
\newcommand\JSM[3]{{\sl J. Soviet Math.} {\bf #1} (#2) #3}
\newcommand\MPLA[3]{{\sl Mod. Phys. Lett.} {\bf A#1} (#2) #3}
\newcommand\JETP[3]{{\sl Sov. Phys. JETP} {\bf #1} (#2) #3}
\newcommand\JETPL[3]{{\sl  Sov. Phys. JETP Lett.} {\bf #1} (#2) #3}
\newcommand\PHSA[3]{{\sl Physica} {\bf A#1} (#2) #3}
\newcommand\PHSD[3]{{\sl Physica} {\bf D#1} (#2) #3}
\def\cKP{{\sf cKP}~}
\def\bfs{{\bf s}}
\def\bfts{{\bf {\tilde s}}}
\def\kere{\mbox{\rm Ker (ad $E$)}}
\def\ime{\mbox{\rm Im (ad $E$)}}
\def\qs{Q_{\bfs}}
\def\cgh{{\widehat {\cal G}}}

\begin{titlepage}
\vspace{-1cm}
\noindent
September, 1995 \hfill{IFT-P/041/95}\\
\phantom{bla}
\hfill{UICHEP-TH/95-9}\\
\phantom{bla}
\hfill{hep-th/9509096}
\\
\vskip .3in

\begin{center}

{\large\bf Constrained KP Models}
\end{center}
\begin{center}
{\large\bf as  Integrable Matrix Hierarchies}
\end{center}
\normalsize
\vskip .4in

\begin{center}
{ H. Aratyn\footnotemark
\footnotetext{Work supported in part by U.S. Department of Energy,
contract DE-FG02-84ER40173}}

\par \vskip .1in \noindent
Department of Physics \\
University of Illinois at Chicago\\
845 W. Taylor St.\\
Chicago, Illinois 60607-7059\\
\par \vskip .3in

\end{center}

\begin{center}
L.A. Ferreira{\footnotemark
\footnotetext{Work supported in part by CNPq}},
J.F. Gomes$^{\,2}$ and A.H. Zimerman$^{\,2}$

\par \vskip .1in \noindent
Instituto de F\'{\i}sica Te\'{o}rica-UNESP\\
Rua Pamplona 145\\
01405-900 S\~{a}o Paulo, Brazil
\par \vskip .3in

\end{center}

\begin{center}
{\large {\bf ABSTRACT}}\\
\end{center}
\par \vskip .3in \noindent

We formulate the
constrained KP hierarchy (denoted by \cKP$_{K+1,M}$)
as an affine ${\widehat {sl}} (M+K+1)$ matrix integrable hierarchy
generalizing the Drinfeld-Sokolov hierarchy.

Using an algebraic approach, including
the graded structure of the generalized Drinfeld-Sokolov hierarchy,
we are able to find several new universal
results valid for the \cKP hierarchy.
In particular, our method yields a closed expression for the second bracket
obtained through Dirac reduction of any untwisted affine Kac-Moody current
algebra. An explicit example is given for the case ${\widehat {sl}} (M+K+1)$,
for which a closed expression for the general recursion operator is also
obtained.
We show how isospectral flows are characterized and grouped according to
the semisimple {\em non-regular} element $E$ of $sl (M+K+1)$ and the content
of the center of the kernel of $E$.

\end{titlepage}

\sect{Introduction}
The constrained KP (\cKP) hierarchy occupies one of the central positions in
the current study of integrable hierarchies.
This is mainly due to the fact that it represents a direct
generalization of the KdV models and includes an impressive list
of partial differential soliton equations
\ct{konop,cheng,dickey,oevels,BX9305,ANP,ANP2}.
Also important are the relationships of the \cKP hierarchy to several
physically relevant models (like Toda models and discrete matrix models).

Let us recapitulate the most general form of the Lax operator belonging to
the \cKP hierarchy:
\be
L = D^{K+1} + \sum_{l=0}^{K-1} u_l D^l + \sum_{i=1}^M \Phi_i D^{-1} \Psi_i
\lab{f-5}
\ee
and subjected to the following flow evolution equations:
\be
\partder{ L}{t_n} = \lb \( L^{n/(K+1)} \)_{+} \, , \, L \rb .
\lab{f-5a}
\ee
We will denote the hierarchy defined by \rf{f-5} and \rf{f-5a}
as {\sf cKP}$_{K+1,M}$.
There are several different parametrizations (obtained by acting with
various Miura maps)
of the coefficients
$u_l, \Phi_i,\Psi_i$ in \rf{f-5}, defining various reformulations of
the {\sf cKP}$_{K+1,M}$ hierarchy.
It is, for instance, known that the Lax operator from \rf{f-5}
can be rewritten as a ratio $L = L_{M+K+1} /L_M$ of two purely differential
operators $L_{M+K+1}$ and $L_M$ of orders ${M+K+1}$ and $M$ respectively.

Here we present a different parametrization governed by the
Zakharov-Shabat equation associated with the ${\widehat {sl}} (M+K+1)$ algebra.
So, instead of working with calculus of the pseudo-differential operators, we
work here with the generalized Drinfeld-Sokolov matrix hierarchy
\ct{D-S,wilson,GIH1,GIH2,mcintosh,PGMS} associated in our case with
the semisimple {\em non-regular} element $E$ of ${\hat {sl}} (M+K+1)$.
The outcome of our construction is that to a given $sl (N+1)$ algebra one can
associate various scalar Lax representations of the
{\sf cKP}$_{K+1,M}$ hierarchies with $M+K=N$ and $M,K \geq 1$.
The special case of $K=0,M=N$ has been
treated in \ct{AGZ} and shown to correspond to the generalized NLS hierarchy
\ct{FK83}, which in turn generalizes the AKNS hierarchy \ct{FNR83,ne85,2boson}
for which $K=0,M=1$.

The paper is organized as follows. In Section 2 the connection between
the generic matrix eigenvalue problem we are interested in and the
pseudo-differential Lax operator of the \cKP type is established.
Section 3 provides the algebraic foundation, within the generalized
Drinfeld-Sokolov hierarchy for our model, with subsection 3.1 dealing with
the example of the ${\widehat {sl}} (M+K+1)$ algebra.
Section 4 examines the Zakharov-Shabat equation for the problem and provides
the construction of the recurrence operator. In Section 5 the second bracket of
the {\sf cKP}$_{K+1,M}$ hierarchy is obtained as a Dirac bracket,
where the matrix hierarchy is considered as a constrained system.
We conclude with Section \ref{sec:final} suggesting few possible applications
 and extensions of our results.

\sect{Matrix Eigenvalue Problem and {\sf cKP} Lax Operators}
Consider the matrix eigenvalue problem
\be
L \Psi = (D + A  + \l E ) \Psi =0
\lab{lpsi}
\ee
for the $(M+K+1) \times (M+ K+1)$ Lax matrix operator
$L= D + A  + \l E$  given by:
\be
L= \left(\begin{array}{cccccccccc}
D    & 0  &\cdots &0     &  q_1     & 0      &\cdots& \cdots & \cdots  &0 \\
0      & D&0      &\cdots&  q_2     & 0      &\cdots& \cdots & \cdots &0 \\
\vdots &    &\ddots &      &\vdots    & 0      &\cdots &\cdots & \cdots
&\vdots \\
0      &    &       &D   & q_M      & 0      &\cdots &\cdots & \cdots &0 \\
r_1    & r_2 &\cdots&r_M & D- v_1   &   \l   &  0     &\cdots &\cdots & 0 \\
0      &    & \cdots & 0  & 0  & D-v_2 &    \l & 0 &  \cdots &\vdots \\
0      &    & \cdots & 0  & 0  & 0  & D-v_3 &    \l & \cdots &\vdots \\
\vdots &    & \cdots & 0  & 0  &\cdots &   0    & \ddots &  \ddots &0\\
\vdots &    & \cdots & 0  & 0  &\cdots &   0    & \ddots &  \ddots &\l \\
0      &   &\cdots   & 0  & \l & 0 & \cdots &0 & \cdots &  D-v_{K+1}
\end{array} \right)
\lab{ma}
\ee
and acting in \rf{lpsi} on the $(M+ K+1)$ column $\Psi$ such that
$\Psi^T = \(\psi_1, \psi_2, \ldots, \psi_{M+K+1}\)$.
$D$ is a differential operator which acts on the function $f$ according to
$\sbr{D}{f}=f^{\pr}$.
We impose the condition $ \sum_{i=1}^{K+1} v_i = 0 $.
Similar matrix operators have appeared in e.g. \ct{mcintosh,liu}.

We write explicitly the linear problem \rf{ma} as
\br
\pa \psi_i + q_i \psi_{M+1} &=& 0 \qquad i=1, \ldots, M  \nonu \\
\sum_{i=1}^M r_i \psi_i + \( \pa - v_1 \) \psi_{M +1} + \l \psi_{M+2}
&=& 0 \nonu\\
( \pa - v_r ) \psi_{M+r} + \l \psi_{M+r+1} &=& 0
\qquad r =2, \ldots, K\nonu\\
\l \psi_{M+1} + ( \pa - v_{K+1} ) \psi_{M+K+1} &=& 0
\lab{lineqs}
\er
Equation \rf{lineqs} gives rise to $ K+1$ scalar Lax eigenvalue equations:
\be
L_j \psi_{M+j} = (- \l)^{K+1} \psi_{M+j}  \quad ; \quad j = 1, \ldots, K+1
\lab{ljpsi}
\ee
where the scalar Lax operator is given by ($r=2, \ldots,K+1)$ :
\br
L_r \eq ( D - v_{r-1} ) ( D - v_{r-2} ) \cdots \( D - v_{2} \)
\( D - v_{1} - \sum_{i=1}^M
r_i D^{-1} q_i \)  ( D - v_{K+1} )  \cdots \( D - v_{r} \)
\nonu \\
L_1 \eq ( D - v_{K+1} ) ( D - v_{K} ) \cdots \( D - v_{2} \)
\( D - v_{1} - \sum_{i=1}^M
r_i D^{-1} q_i \)
\lab{lj}
\er
For all $K+1$ values of $j$ the corresponding Lax operator $L_j$ can be cast
in the form of the Lax operator  \rf{f-5} in {\sf cKP}$_{K+1,M}$ hierarchy.
All of the Lax operators  \rf{lj} can be associated with the
one-matrix eigenvalue problem \rf{ma}.
The question therefore arises whether the above reduction of the matrix
eigenvalue problem determines uniquely the scalar Lax operator.
We will now answer this problem of potential ambiguity
by showing the equivalence between all the Lax operators from \rf{lj}
in the sense of the Darboux-B\"{a}cklund (DB) symmetry.

The following similarity transformations connect the neighboring
Lax operators from \rf{lj}:
\be
\(D - v_{j-1}\)^{-1} L_j \(D - v_{j-1}\) =
T_{j-1}^{-1} L_j T_{j-1} = L_{j-1} \qquad j \geq 2
\lab{tjlj}
\ee
where we have introduced the operator $T_j = \P_j D \P_j^{-1}$
with $ \P_j \equiv \exp ( \int v_{j})$ to emphasize the Darboux-B\"{a}cklund
character of the similarity transformation in \rf{tjlj}.
In addition we have the following eigenvalue equation holding for each
Lax operator $L_j$:
\be
L_{j} \P_{j} =  0 \qquad j = 2, \ldots, K+1
\lab{ljpj}
\ee
Assume now that $L_{j}$ satisfies the Lax flow equation \rf{f-5a}.
Applying it to the equation \rf{ljpj} we find that
$ \( \pa_{t_n} - \(L^{n/(K+1)}_j\)_{+}  \) \P_{j} $ is annihilated
by $L_j$ and we therefore expect that, without loosing generality
to have the following identity
$\( \pa_{t_n} - \(L^{n/(K+1)}_j\)_{+}  \) \P_{j} = \a (t_2, t_3,\ldots)
\P_{j}$,
where the proportionality coefficient $\a$ depends only on times
$t_i $ with $i \geq 2$. Comparing both sides of this identity
we find that $\a=0$ and therefore $\P_{j}$ is an eigenfunction of
$L_{j}$ meaning that:
\be
 \pa_{t_n} e^{ ( \int v_{j})} =
 \(L^{n/(K+1)}_j\)_{+}  e^{( \int v_{j})}
\lab{eigenp}
\ee
Recall now that the DB transformation
$ L \to  T L T^{-1}$, where $T = \Phi D \Phi^{-1}$
with an eigenfunction $\Phi$
preserves the form of the Lax equation \rf{f-5a} i.e.
the DB transformed Lax operator satisfies the same evolution
equation as the original Lax operator (see e.g. \ct{oevela,ANP,ANP2}).
Since
$ L_j  = T_{j-1} L_{j-1} T_{j-1}^{-1}$
and we have equation \rf{eigenp}, we conclude that all the Lax operators
from \rf{lj} are equivalent belonging to the same ``multiplet''
from the DB symmetry point of view.

\sect{Construction of Hierarchies}

In this section we provide the basic ingredients for the construction of the
type of integrable hierarchies we are going to consider. The discussion is
based on the method of references \ct{D-S,wilson,GIH1,GIH2,mcintosh}.

Let $\cgh$ be an affine Kac-Moody algebra, and  $\lie$ be the
 finite dimensional simple Lie algebra associated to it.
The integral gradations of $\cgh$ are defined by vectors
$\bfs = \( s_0,s_1, \ldots, s_r\)$ \ct{kac}, where $s_i$ are non
negative  relatively  prime integers, and $r\equiv {\rm rank}\, \lie$.
The corresponding grading operator is given by
\be
\qs \equiv \sum_{a=1}^{r} s_a\, {2 \l_a \cdot H^0\over \a_a^2}
+ N_{\bfs} d
\lab{grading}
\ee
where $H^0_a$, $a=1,2,\ldots, r$, are the Cartan subalgebra generators
of $\lie$, $\l_a$ its fundamental weights satisfying
${2 \l_a \cdot \a_b \over \a_b^2}= 2 \d_{ab}$, with $\a_a$ being the
simple roots of $\lie$.
$d$ is the usual derivation of $\cgh$, responsible for
the homogeneous gradation of $\cgh$, corresponding to
$\bfs_{\rm hom}= (1,0,0,\ldots ,0)$. In addition, we have,
$N_{\bfs} \equiv \sum_{i=0}^{r} s_i m_i^{\psi}$,
$\psi = \sum_{a=1}^{r}  m_a^{\psi} \a_a$, $m_0^{\psi} = 1$,
where $\psi$ is the highest positive root of $\lie$. So, we have
\be
\cgh = \bigoplus_{n\in \IZ} \cgh_n(\bfs )\, , \qquad
\lb \qs \, , \, \cgh_n(\bfs )\rb = n\, \cgh_n(\bfs )  \, , \qquad
\lb \cgh_m(\bfs ) \, , \, \cgh_n(\bfs )\rb \subset \cgh_{m+n}(\bfs )
\ee

Introduce the Lax matrix operator
\be
L \equiv \pa_x + E + A
\lab{laxop}
\ee
where $E$ is a semisimple element of $\cgh$, lying in $\cgh_1$, and $A$ is a
potential belonging to the subalgebra $\cgh_0$. The construction works
equally well with $E$ belonging to any subspace $\cgh_n$, $n>0$, and $A$
having grade components ranging from $0$ to $n-1$.
However, such general setting will not be needed in what follows.

The fact that $E$ is semisimple means that $\cgh$ can be decomposed into the
kernel and image of the adjoint action of $E$
\be
\cgh = \kere + \ime
\ee
As a consequence of Jacobi identity one has
\be
\lb \kere \, , \, \kere \rb \subset \kere \, , \qquad
\lb \kere \, , \, \ime \rb \subset \ime
\ee

Using the fact that $E$ is semisimple, one can perform a
gauge transformation to rotate the Lax operator into $\kere$.
Consider
\be
L_0 \equiv U \, L \, U^{-1}
\equiv \pa_x + E + \sum_{j=-\infty}^{0} K^{(j)}
\equiv \pa_x + E + K_0
\lab{rotate}
\ee
where $U$ is an exponentiation of negative grade generators,
$U = \exp {\sum_{j=1}^{\infty} T^{(-j)}}$, with
$T^{(-j)} \in \cgh_{-j}(\bfs )$. Decomposing \rf{rotate} into $\qs$
eigensubspaces,
we get equations of the form
$K^{(j)} = - \lb E \, , \, T^{(j-1)}\rb + X^{(j)}$,
where $X^{(j)}$ depends on $T^{(m)}$'s for $m> j-1$. Therefore,
starting from the highest grade component, one can choose $T^{(j-1)}$
to exactly cancel the component of $X^{(j)}$ in $\ime$. Consequently,
one gets  $K^{(j)}\in \kere$.
Notice that the choice of $T^{(j-1)}$ is not unique, since its
component in $\kere$ is not relevant in the cancellation of the
$\ime$ component of $X^{(j)}$. In addition, $T^{(j-1)}$ is
determined  as a local polynomial of the
components of the potential $A$ and its $x$-derivatives.

The flow equations for the hierarchies are constructed in a quite
simple way  \ct{wilson,GIH1}. Consider a constant element
$b^{(N)}$, with grade $N$ ($N>0$), belonging to the center
 of $\kere$. Then, from the considerations above one gets that
$b^{(N)}$  commutes with $L_0$, and so, from \rf{rotate} one has
\be
\lb L \, , \, U^{-1}\, b^{(N)}\, U  \rb =0
\lab{bndef}
\ee
or
\be
\lb L \, , \, \( U^{-1}\, b^{(N)}\, U \)_{\geq 0}\rb =
-  \lb L \, , \, \( U^{-1}\, b^{(N)}\, U \)_{< 0}\rb
\lab{work}
\ee
where $\( \cdot \)_{\geq 0}$ and $\( \cdot \)_{<0}$ mean non negative and
negative grade components respectively.
One observes that the l.h.s. of \rf{work} has components with
grades varying from $0$ to $N+1$, and the r.h.s.  has
grades varying from  $-\infty$ to $0$. Consequently, both sides
of \rf{work} have to lie on the subalgebra $\cgh_0$.
It is therefore consistent to introduce, for each element
$b^{(N)}$ at the center of $\kere$, with grade $N$ ($N>0$)
a flow equation as
\be
{d\, L \o {d\, t_{b^{(N)}}}} ={d\, A \o {d\, t_{b^{(N)}}}} \equiv
\lb L \, , \, B_{b^{(N)}} \rb
\lab{flow}
\ee
where
\be
B_{b^{(N)}} \equiv \( U^{-1}\, b^{(N)}\, U \)_{\geq 0} \equiv
\sum_{j=0}^{N} B_{b^{(N)}}^{(j)} \, ,  \qquad  B_{b^{(N)}}^{(j)} \in
\cgh_j(\bfs )
\lab{bn}
\ee
Notice that $B_{b^{(N)}}$ is a polynomial of the components of $A$ and
its $x$-derivatives.

{}From \rf{rotate} and \rf{flow} one gets
\be
{d\, L_0 \o {d\, t_{b^{(N)}}}} =
\lb L_0 \, , \, {\tilde B}_{b^{(N)}} \rb \, , \quad {\rm with} \quad
{\tilde B}_{b^{(N)}} \equiv U\, B_{b^{(N)}}\, U^{-1} +
{{d\, U}\o {d\, t_{b^{(N)}}}}U^{-1}
\lab{flow0}
\ee
In fact, ${\tilde B}_{b^{(N)}}$ lies in $\kere$. In order
to see that, we denote
${\tilde B}_{b^{(N)}} = {\tilde B}_{b^{(N)}}^{K} + {\tilde B}_{b^{(N)}}^{I}$,
with
${\tilde B}_{b^{(N)}}^{I} \in \ime$ and ${\tilde B}_{b^{(N)}}^{K} \in \kere$.
Then, splitting \rf{flow0} in its $\kere$ and $\ime$
components one gets
\be
{d\, K_0 \o {d\, t_{b^{(N)}}}} - \pa_x \, {\tilde B}_{b^{(N)}}^{K}
= \lb K_0\, , \, {\tilde B}_{b^{(N)}}^{K} \rb
\lab{flowker}
\ee
and
\be
\pa_x \, {\tilde B}_{b^{(N)}}^{I} + \lb E + K_0 \, , \,
{\tilde B}_{b^{(N)}}^{I} \rb = 0
\lab{flowim}
\ee
The highest grade component of \rf{flowim} is
$\lb E  \, , \, \({\tilde B}_{b^{(N)}}^{I}\)_{N} \rb =0$, with
$\({\tilde B}_{b^{(N)}}^{I}\)_{N} \equiv {\tilde B}_{b^{(N)}}^{I}
\cap \cgh_{N}(\bfs )$.
Since there is no intersection between $\kere$ and $\ime$, it
follows that $\({\tilde B}_{b^{(N)}}^{I}\)_{N} = 0$.
Following this reasoning one concludes that
${\tilde B}_{b^{(N)}}^{I} =0$, and so ${\tilde B}_{b^{(N)}}$ given in
\rf{flow0} lies in $\kere$.

Notice that if $\kere$ is abelian (as is a case when $E$ is regular),
then \rf{flowker} constitutes a local conservation law.

The flows defined in \rf{flow} commute, as a consequence of the fact that
${\tilde B}_{b^{(N)}} \in \kere$, and that $b^{(N)}$ belongs to the center of
$\kere$. Indeed, those facts imply that $\lb {d\,\,\,\,\o {d\, t_{b^{(N)}}}} -
{\tilde B}_{b^{(N)}}\, , \, b^{(M)}\rb = 0$. Conjugating with $U$, one gets
${d\,\,\,\,\o {d\, t_{b^{(N)}}}}\( U^{-1}b^{(M)} U\) = \lb B_{b^{(N)}} \, , \,
U^{-1}b^{(M)} U\rb$.
Taking the positive grade part and subtracting the same
 relation with $b^{(M)}$ and $b^{(N)}$ interchanged, one gets
\be
{d\, B_{b^{(M)}}  \o {d\, t_{b^{(N)}}}} - {d\, B_{b^{(N)} }
      \o {d\, t_{b^{(M)}}}}
= \lb B_{b^{(N)}} \, , \, U^{-1}b^{(M)} U\rb_{\geq 0} -
\lb B_{b^{(M)}} \, , \, U^{-1}b^{(N)} U\rb_{\geq 0}
\ee
But
$\lb B_{b^{(N)}} \, , \, U^{-1}b^{(M)} U\rb_{\geq 0} = \lb B_{b^{(N)}} \, , \,
B_{b^{(M)}}\rb + \lb B_{b^{(N)}} \, , \,
\( U^{-1}b^{(M)} U\)_{<0}\rb_{\geq 0}$.
Since $b^{(N)}$ and $b^{(M)}$ commute, it follows that
$    \lb \( U^{-1}b^{(M)} U\)_{\geq 0}\, , \, U^{-1}b^{(N)} U
\rb = - \lb \( U^{-1}b^{(M)} U\)_{<0}\, , \, U^{-1}b^{(N)} U\rb$.
    Taking the positive grade part of it one gets
$\lb B_{b^{(M)}} \, , \, U^{-1}b^{(N)} U\rb_{\geq 0} =
\lb B_{b^{(N)}} \, , \, \( U^{-1}b^{(M)} U\)_{<0} \rb_{\geq 0}$.
    Therefore one concludes that
\be
{d\, B_{b^{(M)}}  \o {d\, t_{b^{(N)}}}} - {d\, B_{b^{(N)}}
  \o {d\, t_{b^{(M)}}}}
+ \lb B_{b^{(M)}} \, , \, B_{b^{(N)}}\rb = 0
\ee
and due to eq. \rf{flow} that is a sufficient condition for the
flows to commute:
\be
\lb {d\, \,   \o {d\, t_{b^{(M)}}}} \, , \,
{d\, \,   \o {d\, t_{b^{(N)}}}}\rb \, L =0
\ee

Notice that the gauge transformations
\be
L \ra e^{S}\, L \, e^{-S}\, , \qquad
B_{b^{(N)}} \ra e^{S}\, B_{b^{(N)}} \, e^{-S}\, , \qquad
{d\, S \o {d\, t_{b^{(N)}}}} =0 \, , \qquad
S \in \cK \equiv \cgh_0 \cap \kere
\lab{gauge}
\ee
are symmetries of the flows equations \rf{flow}, in the sense that they
preserve the form of the Lax operator $L$. Associated to such symmetries
we have conserved quantities. Indeed, the component of zero grade on the
r.h.s. of
\rf{work} is $\lb E \, , \, \( U^{-1}\, b^{(N)}\, U \)_{-1}\rb$. But that
implies that the l.h.s. of \rf{work}, and consequently both sides of \rf{flow},
have no components in $\kere$. Then
\be
{d\,  A^{\cK}    \o {d\, t_{b^{(N)}}}}=0 \, , \qquad
A^{\cK} \equiv A \cap \kere
\lab{aker}
\ee

Therefore, if we choose $A^{\cK}=0$ at $t_{b^{(N)}} =0$,  it will
remain zero for all times. That is a reduction procedure which we will use
below to  obtain the constrained KP hierarchies from the above formalism.
We shall decompose the potential $A \in \cgh_0$ as
\be
A \equiv A_0 + A^{\cK}
\lab{asplit}
\ee
with $A^{\cK} \in \cK$, and $A_0$ lying in the complement $\cM$ of $\cK$ in
$\cgh_0$.
$A_0$ contains therefore the dynamical variables of the integrable hierarchy.

Since we are working with loop algebras (vanishing central term) it is useful
to work with finite matrix representations.  The commutation relations for
$\cgh$ can be written as
\be
\sbr{T_a^m}{T_b^n} = f_{ab}^c T_c^{m+n} \, \qquad
\sbr{d}{T_a^m}=m\, T_a^m
\ee
where $T_a^0\equiv T_a$, $a=1,2,\ldots, {\rm dim}\, \lie$, are the generators
of the finite simple Lie algebra $\lie$, and $f_{ab}^c$ are its structure
constants. If one has a (finite) matrix representation of $\lie$ then one can
construct a representation of $\cgh$ by replacing
\be
T_a^m \ra z^m \, T_a
\lab{rep1}
\ee
where $z$ is a complex parameter. However, in some calculations we will be
interested in another representation of such type, where the powers of
the complex parameter count the grade w.r.t. $\qs$ defined in \rf{grading}.
Accordingly we replace
\be
T_a^m \ra \l^l \, T_a \, \qquad l = g_a + m N_{\bfs}
\lab{rep2}
\ee
where $\l$ is a complex parameter, and
\be
\sbr{\sum_{b=1}^{r} s_b\, {2 \l_b \cdot H^0\over \a_b^2}}{T_a}= g_a\, T_a
\ee
Notice that $g_a$ take values between $-N_{\bfs}+1$ and $N_{\bfs}-1$.

In the representation \rf{rep1} one has $d \equiv z\,{d\,\o{dz}}$.
Therefore if $\sbr{\qs}{X}= x_{\bfs} X$, with $\qs$ given by \rf{grading} one
has $\sbr{\qs}{z\, X}= (x_{\bfs}+ N_{\bfs}) z\, X$. Now, if $b^{(N)}$ is an
element of the center of $\kere$, so is $z\, b^{(N)}$. We shall denote
$b^{(N+N_{\bfs})} \equiv  z\, b^{(N)}$. Therefore, $z\,U^{-1}b^{(N)}U =
U^{-1}b^{(N+N_{\bfs})}U$, and so
\be
z\,\( U^{-1}b^{(N)}U\)_{\geq 0} + z\,\( U^{-1}b^{(N)}U\)_{<0} =
\( U^{-1}b^{(N+N_{\bfs})}U\)_{\geq 0} + \( U^{-1}b^{(N+N_{\bfs})}U\)_{<0}
\lab{cong1}
\ee
where $\( \cdot\)_{\geq 0}$ and $\( \cdot\)_{<0}$ mean the non negative
and negative  $\bfs$-grade components respectively. But since multiplication
by $z$ increases the $\bfs$-grade by $N_{\bfs}$ we have that the positive
part of \rf{cong1} leads to
\be
B_{b^{(N+N_{\bfs})}} = z\, B_{b^{(N)}} + z\, \sum_{j=1}^{N_{\bfs}}
\( U^{-1}b^{(N)}U\)_{-j}
\lab{cong2}
\ee
Notice the second term on the r.h.s. of \rf{cong2} have components
with $\bfs$-grades varying from $0$ to $N_{\bfs}-1$. But, analyzing
\rf{grading}, one concludes that any generator having positive grade w.r.t.
$d$, has necessarily $\bfs$-grade greater or equal to $N_{\bfs}$. Therefore,
we conclude that the quantity $z\, \sum_{j=1}^{N_{\bfs}} \(
U^{-1}b^{(N)}U\)_{-j}$ is $z$ independent in the representation \rf{rep1}.

\subsection{The case of ${\bf {\widehat {sl}}(M+K+1)}$}
\label{slmk}

We now apply the above formalism
to the example of the affine Kac-Moody algebra
$\cgh = {\widehat {sl}} \( M+K+1 \)$, ($A_{M+K}^{(1)}$)
without a central term (i.e. a loop algebra), furnished
with the following gradation $\bfs$ and corresponding grading operator $\qs$:
\be
\bfs = ( 1, \underbrace{0, \ldots ,0}_{M}, \underbrace{1, \ldots,1}_{K}\, )
\quad ; \quad \qs = \sum_{j=M+1}^{M+K} \l_j \cdot H^{(0)} + (K+1) d
\lab{a1}
\ee
We will denote the simple roots of ${\widehat{sl}} \( M+K+1 \)$ by $\a_j$,
$j=0,1,\ldots, M+K$, with $\a_0 \equiv -\psi$, and $\psi$ being the highest
positive root of $\lie = sl \( M+K+1 \)$, which is the subalgebra of
${\widehat{sl}} \( M+K+1 \)$ commuting with $d$. Since
${\widehat{sl}} \( M+K+1 \)$ is simply laced we normalize the roots such
that $\a_j^2=2$.
The ordering of the roots is such that for $j\neq k$, $\a_j \cdot \a_k =
-\d_{j,k\pm 1\,({\rm mod}\, M+K+1)}$, $j,k=0,1,\ldots, M+K$.
The fundamental weights $\l_j$ satisfy $\l_j\cdot \a_k = \d_{j,k}$.
We choose the semisimple element $E$ to be
\be
E = \sum_{j=M+1}^{M+K}  E^{(0)}_{\a_{j}}
+  E^{(1)}_{-(\a_{M+1}+ \cdots+\a_{M+K})}
\lab{a2}
\ee
One can check that each generator in $E$ has grade one w.r.t. $\qs$.
Hence
\be
\sbr{\qs}{E} = E
\lab{a5}
\ee
The zero grade subalgebra is:
\be
\cgh_0 \equiv \{ \lie_0^{(0)} \equiv sl (M+1) \quad ; \quad \a_j \cdot H^{(0)}
\quad, \quad j = M+1, M+2, \ldots, M+K \} \lab{a6}
\lab{a7}
\ee
where $\lie_0^{(0)}$ is the  $sl (M+1)$ subalgebra of $\lie=sl (M+K+1)$, with
simple roots   $\a_1, \ldots, \a_M$.

For $E$ defined in \rf{a2} we have
\be
\kere = \{ {\hat K}_0 \equiv {\widehat{sl}} (M) \oplus {\hat U} (1) \; , \;
{\hat {\cH}}_K \}
\lab{a8}
\ee
where ${\widehat{sl}} (M)$ is the affine Kac-Moody subalgebra of
$\cgh = {\widehat{sl}} (M+K+1)$ with simple roots  $\a_j$, $j=1, 2, \ldots,
M-1$ and $\a_0=-(\a_1 + \a_2 + \ldots + \a_{M-1})$. The Kac-Moody algebra
${\hat U} (1)$ is generated by $\l_M \cdot H^{(k)}$, $k\in \IZ$.
In addition ${\hat{\cH}}_K$ is the principal Heisenberg subalgebra of
${\widehat{sl}} (K+1) \in {\widehat{sl}} (M+K+1)  $ containing $E$.
We can denote its generators by $E_{N}$, where $N$ contains the integers
$1,2,3, \ldots, K$ (modulo $K+1$) or in other words, the integers without
the multiples of $(K+1)$.
In this notation we have
\br
E_{l+(K+1)n} \eq E^{(n)}_{\a_{M+1}+\a_{M+2}+\ldots +\a_{M+l}} +
E^{(n)}_{\a_{M+2}+\a_{M+3}+\ldots + \a_{M+l+1}} + \ldots
\nonu \\
&+&
E^{(n)}_{\a_{M+K-l+1} + \a_{M+K-l+2} +\ldots +\a_{M+K-1}+\a_{M+K}} \nonu \\
&+&
E^{(n+1)}_{-(\a_{M+1}+\a_{M+2}+\ldots +\a_{M+K-l+1})} +
E^{(n+1)}_{-(\a_{M+2}+\a_{M+3}+\ldots +\a_{M+K-l})} + \ldots
\nonu \\
&+&
E^{(n+1)}_{-(\a_{M+l} + \a_{M+3}+\ldots +\a_{M+K})}
\er
with $l=1, 2, 3, \ldots,K$, and so
\be
\sbr{\qs}{E_{l+(K+1)n}} = (l+(K+1)n)\, E_{l+(K+1)n}
\ee
Notice that $E_1 \equiv E$. In addition, since we are
working here with the case of loop algebra ($c=0$):
\be
\sbr{E_{N}}{E_{N^{\pr}}} = 0
\ee

In fact, in the representation \rf{rep2} of the loop algebra one has
\be
E = \l \, {\tilde E} = \l \, \( \sum_{j=M+1}^{M+K}  E_{\a_{j}}
+  E_{-(\a_{M+1}+ \cdots+\a_{M+K})}\)
\lab{ele}
\ee
and in the defining representation of $sl (M+K+1)$, one has
\be
{\tilde E} = \fourmat{0}{0}{0}{{\ti e}}\quad;\quad
{\ti e} =\left(\begin{array}{ccccc}
0&1& & & \\
&0&1& & \\
& &\ddots&\ddots& \\
& & &0&1\\
1& & & &0  \end{array} \right)
\lab{etilde}
\ee
where ${\tilde E}$ and ${\ti e}$ are $(M+K+1) \times (M+ K+1)$ and
$(K+1) \times (K+1)$ matrices, respectively.
Elements of this type are among the generators of the non-equivalent
Heisenberg subalgebras, see e.g. Appendix of \ct{PGMS}  for the cases
of $sl(3)$ and $sl(4)$.

In addition, one has
\be
E_{l+(K+1)n} = \l^{l+(K+1)n} \, \({\tilde E}\)^{l}
\ee
Also for $c=0$ we have
\be
{\rm center }\, \kere = \{  {\hat U} (1) \; , \;{\hat{\cH}}_K \}
\lab{a13}
\ee
where ${\hat U} (1) $ is generated by $\l_M \cdot H^{(k)}\, , \, k \in \IZ$.
Notice that
$\sbr{\qs}{\l_M \cdot H^{(k)}} = k (K+1) \l_M \cdot H^{(k)}$.
Therefore the center of $\kere$ while having
generators of arbitrary grade, has one and only one generator of a given
grade.
Then the choices we have for the elements $b^{(N)}$, introduced in
\rf{bndef}, are
\br
b^{(N)} &=&  E_N \qquad N =1,2, \ldots, K\,\, {\rm mod}\, (K+1) \lab{a16a} \\
b^{(k(K+1))} &=& \l_M \cdot H^{(k)}\, , \, \qquad k \in \IZ
\lab{a16b}
\er
According to \rf{flow} each of the generators from the center of \kere
appearing in \rf{a16a}-\rf{a16b}
will give rise to the corresponding flow with
times $t_{b^{(N)}}, t_{b^{(k(K+1))}}$.
In particular the element $E_1 \equiv E$
will generate the flow corresponding to $\pa / \pa t_1 = \pa / \pa x$.

The gauge symmetries of the model are then given by the transformations
\rf{gauge}, where $S$ belongs to the subalgebra
\be
\cK \equiv  \cgh_0 \cap \kere = \{ sl (M) , \l_M \cdot H^{(0)} \}
\lab{a13a}
\ee
where $sl (M)$ is the subalgebra of $\lie=sl (M+K+1)$ with simple roots
$\a_1 , \a_2 , \ldots, \a_{M-1}$.

The generators of the complement $\cM$ of $\cK$ in $\cgh_0$ are
\be
\cM = \{ P_{\pm i} = E_{\pm (\a_{i} + \a_{i+1} + \ldots +\a_{M})}^{(0)}\, ,
\,\, i=1, 2, \ldots, M, \,\, {\rm and} \,\,\a_a \cdot H^{(0)},
\,\, a=M+1, M+2, \ldots, M+K\}
\lab{pis}
\ee
We then parametrize $A_0$, defined in \rf{asplit}, as follows
\be
A_0  = \sum_{i=1}^{M} \( q_i P_i + r_i {P}_{-i} \) +
\sum_{a=M+1}^{M+K} U_a ( \a_a \cdot H^{(0)})
\lab{a20}
\ee
where $q_i$, $r_i$ and $U_a$ are fields of the model.

As we have shown in \rf{aker}, $A^{\cK}$ is constant in time. Therefore,
we will work with the submodel defined by
\be
A^{\cK} = 0
\ee

The flow equations \rf{flow}, in this case, become
\be
{d\, A_0 \o {d\, t_{b^{(N)}}}} - \pa_x B_{b^{(N)}}^0 =
\lb E+A_0 \, , \, B_{b^{(N)}}^0 \rb
\ee
where $B_{b^{(N)}}^0$ is the constrained $B_{b^{(N)}}$, i.e.
\be
B_{b^{(N)}}^0 = B_{b^{(N)}}\mid_{A^{\cK}=0}
\ee

The effective potential of our submodel lies therefore, on the tangent plane
of the coset space
$\cgh_0/\cK \equiv (sl (M+1) \oplus U(1)^K)/(sl (M) \oplus U(1)_M)$.
 $U(1)_M$ is generated by $\l_M\cdot H^{(0)}$, and consequently is
a linear combination of the generators of the Cartan subalgebra of $sl (M+1)$
and also of all the generators of $U(1)^K$. Remember that $\l_i = K_{ij}^{-1}
\, \a_j$, and the inverse of the Cartan matrix $K_{ij}^{-1}$ of $A_n$ has no
vanishing elements. Those facts prevent $\cgh_0/\cK$ from being a symmetric
space. Indeed, one can verify that
\br
\sbr{P_j}{P_{-j}} &=& \a_M \cdot H^{(0)} + \sum_{i=j}^{M-1}\a_i \cdot H^{(0)}
\nonu\\
&\in& \cK + \cM \, , \qquad P_j , P_{-j} \in \cM
\er
since $ \a_M \cdot H^{(0)}$ have components on both $\cK$ and $\cM$.

However, one has
\br
\sbr{P_j}{P_{-k}} \in \cK \, \quad \mbox{\rm for $j\neq k$} \, , \quad
\sbr{P_j}{P_{k}} = \sbr{P_{-j}}{P_{-k}} = 0 \,
\quad \mbox{\rm for any $j, k$}
\er

\subsubsection{The case $K=0$}

In this case we have $\cgh = {\widehat {sl}} (M+1)$.
The relevant gradation is the homogeneous one, $\bfs = (1,0,0,\ldots, 0)$,
and so $\qs \equiv d$. The semisimple element $E$ is now given by
\be
E = \l_M \cdot H^{(1)}\, , \qquad \sbr{d}{E}=E
\ee
This example was discussed in detail in ref. \ct{AGZ}, and here, we just give a
brief description of it to make contact with the model discussed above.

The grade zero subalgebra is
\be
\cgh_0 = \{ \lie \equiv sl (M+1)\}
\ee
and
\be
\kere = \{ {\widehat {sl}} (M) \oplus {\hat U}(1) \}
\ee
with ${\hat U}(1)$ being generated by $\l_M \cdot H^{(k)}$, $k\in \IZ$.

The center of $\kere$ is just the homogeneous Heisenberg subalgebra of
$\cgh = {\widehat {sl}} (M+1)$, namely
\be
{\rm center}\, \kere = \{ \l_i \cdot H^{(k)}\,\, , \,\,  i=1,2,\ldots,M\,\,
,  \,\, k \in \IZ \}
\ee
Therefore, we can introduce a flow for each element (see \rf{bndef})
\be
b^{(k)}_i \equiv \l_i \cdot H^{(k)} \, , \, \qquad \mbox{\rm $k$ being a
positive integer}
\ee

The gauge subalgebra $\cK$ introduced in \rf{gauge} is
\be
\cK = \{ sl (M) \oplus U(1) \}
\ee
where $sl (M)$ is the subalgebra of $\lie = sl (M+1)$ with simple roots
$\a_1,\a_2 ,\ldots, \a_{M-1}$, and $U(1)$ is generated by $\l_M \cdot
H^{(0))}$. We write the potential $A \in \cgh_0$ as
$A = A^{\cK} + A_0$, with $A^{\cK} \in \cK$ and $A_0$ lying in the
complement $\cM$ of $\cK$ in $\cgh_0$. Then, we parametrize $A_0$ as
\be
A_0  = \sum_{i=1}^{M} \( q_i P_i + r_i {P}_{-i} \)
\ee
where $P_{\pm i}$ were introduced in \rf{pis}. Comparing with
\rf{a20}, we notice an absence of $U_a$'s fields in this special example.
Also $\cgh_0/\cK$ is now a symmetric space.

\sect{Zakharov-Shabat Equation and Recursion Operator}
Recall that within the models considered here the semisimple
element $E$ is given for $\lie = sl (M+K+1)$ by \rf{ele}.
We first notice that $E$ commutes with its conjugated counterpart
$E^{\dagger}$ and therefore, although not Hermitian, may be diagonalized
\ct{OT}.
As a consequence, the Lie algebra $\lie$ under consideration can be
decomposed into graded subspaces, i.e.
\be
\lie = \oplus_{s} \lie_s \quad ;\quad \sbr{E}{\lie_s} = s \lie_s
\lab{gradinga}
\ee
where $s$ can be in general a complex number ($s \in \IC$).
This is a crucial property allowing to solve the Zakharov-Shabat (Z-S)
equation in the manner shown below.
Consider namely the Z-S equation
\be
\pa_m A_0 - \pa B_m + \l \lb E \, , \, B_m \rb
+\lb A_0 \, , \, B_m \rb = 0
\lab{zsa}
\ee
where $A_0$ defined in Section 3 lies, as described there, in subspace
orthogonal to $Ker (ad E)$.

Decomposing $B_m  = \sum_s B^{(s)}_m$ and $A_0  = \sum_s A^{(s)}$ into
components according to the gradation defined by $E$ induces a natural
decomposition of the Z-S equation \rf{zsa} into the zero and non-zero
components:
\br
- \pa B^{(0)}_m + \sum_{x+y=0} \lb A^{(x)} \, , \, B^{(y)}_m \rb &=& 0
\lab{zs-0}\\
\pa_m A^{(s)} - \pa B^{(s)}_m + \l s B^{(s)}_m
+ \sum_{x+y=s} \lb A^{(x)} \, , \, B^{(y)}_m \rb &=& 0
\lab{zs-s}
\er
where in the last equation the summation is over $x, y \in \IC$ and includes
$y=0, x=s$.
Equations \rf{zs-0} and \rf{zs-s} contain components of \rf{zsa} in
$\kere$ and $\ime = \oplus_{s \in \IC - \{0\}} \lie_s$.
We now assume the following expansions:
\be
B^{(s)}_m = \sum_{i=0}^{m-1}  B^{(s)}_m (i) \l^i \qquad s \ne 0
\lab{AN}
\ee
for all non-zero gradation components of $B_m$, while for the zero component
we find from \rf{zs-0} by integration:
\be
B^{(0)}_m = D^{-1} \( \sum^{x+y=0}_{x,y \ne 0}
\lb A^{(x)} \, , \, B^{(y)}_m \rb \)  + \l^m \Lambda_m
\lab{B0}
\ee
where the last (integration constant) term on the right hand side of \rf{B0}
is of higher order than those in \rf{AN}.
Its presence is allowed by the structure of \rf{zsa} as long as
$\Lambda_m  \in \kere $.

Inserting \rf{AN} in \rf{zs-s} we find by collecting the coefficients of
$\l^{k-1}$:
\be
l B^{(l)}_m (k-1) = - \pa B^{(l)}_m (k)
- \sum^{x+y=0}_{x,y \ne 0} \lb A^{(l)} \, , \, D^{-1} \lb A^{(x)}\, ,\,
B^{(y)}_m (k) \rb \rb -
\sum^{x+y=l}_{x,y \ne l} \lb A^{(l)} \, , \, B^{(y)}_m (k)  \rb
\lab{Bl}
\ee
for $k=m$ we obtain
\be
l B^{(l)}_m (m-1) = - \lb A^{(l)} \, , \, \Lambda_m  \rb
\lab{Blmone}
\ee
{}From \rf{Bl} we find that the general solution can be rewritten as
\be
B^{(l)}_m (k-1) = \sum_{y \ne 0} \P_{l,y} B^{(y)}_m (k)
\lab{PhiB}
\ee
with
\be
\P_{l,y} = { 1 \o l} \( D \d_{l,y} - \sum_{x \in Grad (\cM)}
\llb  ad_{A^{(l)}} D^{-1} ad_{A^{(x)}}  \d_{x,-y}
+ ad_{A^{(x)}} \d_{l-x,y}      \rrb \)
\lab{Ply}
\ee
Using \rf{PhiB} repeatedly we are led to:
\be
B^{(l)}_m (0) = \sum_{y_1,\ldots,y_{m-1}} \P_{l,y_1}\P_{y_1,y_2} \cdots
\P_{y_{m-2},y_{m-1}} B^{(y_{m-1})}_m (m-1)
\lab{Bl0}
\ee
and after taking into account \rf{Blmone}
\be
B^{(l)}_m (0) = \sum_{y_1,\ldots,y_{m-1}} \P_{l,y_1}\P_{y_1,y_2}\cdots
\P_{y_{m-2},y_{m-1}} {1 \o y_{m-1}} ad_{\Lambda_m}  A^{(y_{m-1})}
\lab{Bl0a}
\ee
Projecting \rf{zs-s} on the $\l$-independent component we find
\be
\pa_m A^{(l)} =  \pa B^{(l)}_m (0)
- \sum_{x+y=l} \lb A^{(x)} \, , \, B^{(y)}_m  (0) \rb
= l \sum_{y} \P_{l,y} B^{(y)}_m (0)
\lab{zs-s0}
\ee
Substituting the solution of $B^{(l)}_m (0)$ in terms of $A^{(y)}$
from \rf{Bl0a} we arrive at:
\be
{1 \o l} \pa_m A^{(l)} = \sum_{y_1,\ldots,y_{m}}
\P_{l,y_1}\P_{y_1,y_2} \cdots \P_{y_{m-1},y_{m}}
  ad_{\Lambda_m} \(  {A^{(y_{m})}\o y_{m}} \)
\lab{recal}
\ee
This expression leads to a recursion operator relating consecutive flows
belonging the same family of flows generated by the specific element
$ \Lambda_m = ({\ti E}^l , \l_M \cdot H^{(0)}) $ from the center of $\kere$
as explained in discussion around equations \rf{a16a}-\rf{a16b}.
Therefore these consecutive times have indices modulo
$K+1$ for the case of $sl (M+K+1)$.

To get the closed expression for the recursion operator we compare
equation \rf{recal} to the corresponding expression for
$\pa_{m-K-1} A^{(l)}$. These flows are related through:
\be
{1 \o l} \pa_m A^{(l)} = \sum_{y_1,\ldots,y_{K+1}}
\P_{l,y_1}\cdots \P_{y_{K},y_{K+1}} \pa_{m-K-1} \( {A^{(y_{K+1})}
  \o y_{K+1}} \) \equiv \cR_{l, K+1} \pa_{m-K-1} A^{(y_{K+1})} /y_{K+1}
\lab{recurop}
\ee
This yields an expression for the recurrence operator $\cR$ as
$\cR = \P^{K+1}$ for the $sl (M+K+1)$-matrix hierarchy.

\sect{The Second Bracket Structure}

The potential $A$ introduced in \rf{laxop} is an element of the subalgebra
$\cgh_0$, and so we shall denote it as
\be
A = \eta^{ab} \vp \( T_a \) T_b \,\; , \qquad
\vp \( T_a  \) \equiv \Tr \( T_a  \, A\)
\lab{51a}
\ee
where $\eta^{ab}$ is the inverse of the trace form of $\cgh_0$,
$\eta_{ab}\equiv \Tr \( T_a\, T_b\)$, $a,b=1,2,\ldots, {\rm dim}\, \cgh_0$.

There is a natural Poisson bracket structure for the manifold spanned by
$\vp$'s components of $A$, induced by the $\lie_0$-KM current algebra:
\be
{\pbr{\vp \( T\) (x)}{\vp \( T^{\pr}\) (y)}}_{PB} =  \vp \( \sbr{T}{T^{\pr}}\)
(x) \d (x-y) + \Tr \( T\,T^{\pr}\) \d^{\pr} (x-y) \, , \;\quad
T\, , \, T^{\pr} \in {\hat {\lie}}_0
\lab{pbkm}
\ee

The model we are interested in, is a constrained system where the components
of $A$ in $\kere$ are set to zero (see \rf{aker}). The bracket structure of
such submodel is then given by the Dirac bracket associated to \rf{pbkm}.
We denote by $\cK_i$, $i=1,2,\ldots, {\rm dim}\, \cK$,  and $M_r$,
$r=1,2,\ldots, {\rm dim}\, {\hat {\lie}}_0 - {\rm dim}\, \cK$, the generators
of the
subalgebra $\cK$, defined in \rf{gauge}, and of its complement $\cM$ in
 ${\hat {\lie}}_0$, respectively.

The Dirac matrix is given by
\be
\Delta_{ij}\( x,y\) \equiv \pbr{\vp\( \cK_i \)\( x\)}{\vp \( \cK_j\)\( y\)}
\approx \eta_{ij} \d^{\pr}\( x-y\)
\lab{diracmatrix}
\ee
where $\eta_{ij} \equiv \Tr \( \cK_i \cK_j\)$, and $\approx$ means equality
after the constraints are imposed. Therefore
\be
\Delta^{-1}_{ij}\( x,y\) \approx \eta^{ij} \pa_x^{-1}\d \( x-y\)
\qquad i,j=1,2,\ldots, {\rm dim}\, \cK
\lab{invdirac}
\ee

Consequently the Dirac bracket is
\br
\pbr{\vp\( M_r\)\( x\)}{\vp\( M_s\)\( y\)}_{DB} &=&
\vp \( \sbr{M_r}{M_s}\)\( x\) \d\( x-y\) + \Tr \( M_r M_s\) \d^{\pr}\( x-y\)
\nonu\\
&+&  \eta^{ij}\vp\( \sbr{\cK_i}{M_{r}}\)\( x\)
\vp\( \sbr{\cK_j}{M_{s}}\)\( y\) \pa_x^{-1}\d \( x-y\)
\lab{dirac}
\er

The subspace $\cM$ constitutes a representation of the subalgebra $\cK$,
\be
\sbr{\cK_i}{M_r} = R_{rs}\( \cK_i\) \, M_s
\lab{mrep}
\ee
Therefore, the second term on the r.h.s. of the bracket \rf{dirac} can be
calculated using representation theory. The relevant representation here,
is the tensor product of the representation, $R\otimes R$, defined by the
linear functionals  $\vp\( M_{r}\)\( x\)$. We then write
\br
X_{rs}\( x,y\) &\equiv& \eta^{ij}\vp\( \sbr{\cK_i}{M_{r}}\)\( x\)
\vp\( \sbr{\cK_j}{M_{s}}\)\( y\) \pa_x^{-1}\d \( x-y\)\nonu\\
\eq \eta^{ij} R_{rt}\( \cK_i\)R_{su}\( \cK_j\) \vp\( M_{t}\)\( x\)
\vp\( M_{u}\)\( y\) \pa_x^{-1}\d \( x-y\)\nonu\\
&\equiv&
\IC \, \mid M_r\rangle_x \otimes \mid M_s\rangle_y \,\, \pa_x^{-1}\d \( x-y\)
\lab{diracterm}
\er
where
\be
\IC \equiv \eta^{ij} \cK_i \otimes \cK_j
\lab{fakecasimir}
\ee
and where we have denoted states of the representation $R\otimes R$,  as
$\vp\( M_{r}\)\( x\) \vp\( M_{s}\)\( y\) \equiv$
$\mid M_r\rangle_x \otimes \mid M_s\rangle_y$.
So, the space variables $x$ and $y$ define the left and right
entries, respectively, of the tensor product.

The operator \rf{fakecasimir} commutes with any generator
\be
\sbr{\IC}{1\otimes \cK_i + \cK_i \otimes 1}=0
\ee
and according to Schur's lemma, it is proportional to the identity in each
irreducible component of $R \otimes R$. That fact, simplifies substantially the
evaluation of \rf{diracterm}. Notice that $\IC$ is not the quadratic
Casimir operator in $R\otimes R$. That operator is given by
$\IC_{R\otimes R}\equiv \eta^{ij}
\( 1\otimes \cK_i + \cK_i \otimes 1\)\( 1 \otimes \cK_j + \cK_j \otimes 1\)$,
and therefore we have
\be
\IC = \h \( \IC_{R\otimes R} - 1\otimes C - C \otimes 1\)
\lab{casimirrel}
\ee
where $C = \eta^{ij} \cK_i \cK_j$, is the quadratic Casimir in $R$.

Decomposing the representation $R \otimes R$ in its irreducible components, one
can evaluate \rf{diracterm}, using \rf{casimirrel} and the fact that the value
of the quadratic Casimir  operator in an irreducible representation  is
$\l \(  \l + 2 \d\)$, where $\l$ is the highest weight, and
$\d = \h \sum_{\a >0} \a = \sum_{a=1}^{{\rm rank}} \l_a$, with $\a$
being the positive roots, and $\l_a$ the fundamental weights of $\cK$.

\subsection{The case of  ${\bf {\widehat {sl}} (M+K+1)}$-Matrix
Integrable Hierarchy}

We now consider the example of the affine Kac-Moody algebra $sl (M+K+1)$
discussed in subsection \ref{slmk}. The subalgebra $\cK$, and subspace $\cM$
are defined in \rf{a13a} and \rf{pis} respectively. We shall denote by
${\tilde {\cK}}_i$, $i=1,2, \ldots , M^2-1$, the generators of the subalgebra
$sl (M)$ of $\cK$.

One can easily verify that $P_j$ and $P_{-j}$, $j=1,2,\ldots, M$ transform
under the representations $M$ and ${\bar M}$ of $sl (M)$ respectively.
The highest weights and highest weight states of these representations are
$\(\l_1 , P_1\)$ and $\( \l_{M-1}, P_{-M}\)$, respectively.
The remaining generators of $\cM$ \rf{pis}, namely
$\a_a\cdot H^{(0)}$, $a=M+1, M=2, \ldots, M+K$, are scalars under $sl (M)$.
The charges of the $U(1)$ factor of $\cK$
\rf{a13a}, generated by $\l_M \cdot H^{(0)}$, are $1$ for $P_j$, $-1$ for
$P_{-j}$, and $0$ for $\a_a\cdot H^{(0)}$.
Therefore, the representation $R$ of $\cK = sl (M) \oplus U(1)$, defined by
$\cM$ \rf{mrep} is  $R= \( M,1\) + \( {\bar M},-1\) + \( 0,0\)^K$.

We denote the operator \rf{fakecasimir} as
\br
\IC &=& {\tilde \eta}^{ij} {\tilde {\cK}}_i \otimes {\tilde {\cK}}_j
\,\,\, + \,\,\,
{1 \o{\Tr\(\l_M\cdot H^{(0)}\)^2}} \,\,  \l_M\cdot H^{(0)}\otimes \l_M\cdot
H^{(0)}\nonu\\
&\equiv& {\tilde {\IC}}\,\,\, +\,\,\, {{M+K+1}\o{M(K+1)}}\,\, \l_M\cdot
H^{(0)}\otimes \l_M\cdot H^{(0)}
\lab{casimirslmu1}
\er
where ${\tilde \eta}^{ij}$ is the inverse of ${\tilde \eta}_{ij} = \Tr \(
 {\tilde
{\cK}}_i\, {\tilde {\cK}}_j\)$. Notice that $\Tr \( \cdot \, \cdot \)$, as
introduced in  \rf{51a}, is the trace form of the subalgebra $\cgh_0$.

Let us then analyze the various irreducible components of the representation
$R\otimes R$.
The states $\mid P_i \rangle \otimes \mid P_j \rangle$ decompose into the
symmetric and antisymmetric parts, which are the $M(M+1)/2$ and $M(M-1)/2$
irreducible representations of $sl (M)$ respectively. The highest weights of
these representations are $2 \l_1$ and $2 \l_1 - \a_1$ respectively.
Therefore, using \rf{casimirslmu1}, and then \rf{casimirrel} for
${\tilde {\IC}}$, one gets
\br
\IC \, \mid P_i \rangle \otimes \mid P_j \rangle \eq
\( {{M+K+1}\o{M(K+1)}} -  \l_1 \( \l_1 + 2 \d \) \)  \mid P_i \rangle \otimes
\mid P_j \rangle \nonu\\
&+&  \(  \l_1 \(  \l_1 +  \d \)\)\Bigl( \mid P_i \rangle \otimes \mid P_j
\rangle +  \mid P_j \rangle \otimes \mid P_i \rangle \Bigr) \lab{cpipj}\\
&+& {1\o 4}\( \(2 \l_1 - \a_1\) \( 2 \l_1 - \a_1  + 2 \d \)\) \Bigl(
\mid P_i  \rangle
\otimes \mid P_j \rangle  - \mid P_j \rangle \otimes \mid P_i \rangle
\Bigr) \nonu
\er
Denoting the simple roots of $sl (M)$ as $\a_j = e_j - e_{j+1}$,
$j=1,2,\ldots, M-1$, $e_j \cdot e_k = \d_{jk}$, one has
$\l_j = \sum_{k=1}^j e_k - {j\o {M}} \sum_{k=1}^{M} e_k$,
and therefore
$\d = \sum_{k=1}^{M-1} \l_k = \h  \sum_{k=1}^M \( M - 2k + 1\) e_k$.
Consequently
\be
\IC \, \mid P_i \rangle \otimes \mid P_j \rangle =
{1\o{K+1}}\, \mid P_i \rangle \otimes \mid P_j \rangle +
\mid P_j \rangle \otimes \mid P_i \rangle
\lab{symm1}
\ee
By the same arguments, one gets
\be
\IC \, \mid P_{-i} \rangle \otimes \mid P_{-j} \rangle =
{1\o{K+1}}\, \mid P_{-i} \rangle \otimes \mid P_{-j} \rangle +
\mid P_{-j} \rangle \otimes \mid P_{-i} \rangle
\lab{symm2}
\ee

As for the states $\mid P_{i} \rangle \otimes \mid P_{-j} \rangle$, we use the
fact that the tensor product of the $M$ and ${\bar M}$ representations of
$sl (M)$ produces an adjoint and a singlet, i.e.,
$M \otimes {\bar M} = {\rm Adj} +1$. The singlet is the state
$\mid S \rangle \equiv \sum_{j=1}^M \mid P_{j} \rangle \otimes
\mid P_{-j} \rangle$.
The states of the adjoint are $\mid P_{i}
\rangle \otimes \mid P_{-j} \rangle$ for $i\neq j$, and $\mid P_{j} \rangle
\otimes \mid P_{-j} \rangle -  \mid P_{j+1} \rangle \otimes \mid P_{-(j+1)}
\rangle$.
The highest weight of the adjoint is the highest positive root
$\psi = \a_1 + \a_1 + \ldots + \a_{M-1} = e_1 - e_M$. Therefore, using
\rf{casimirslmu1}, and then \rf{casimirrel} for ${\tilde {\IC}}$, one gets
for $i\neq j$
\br
\IC \, \mid P_{i} \rangle \otimes \mid P_{-j} \rangle &=&
 - {{M+K+1}\o{M(K+1)}} \mid P_{i} \rangle \otimes \mid P_{-j} \rangle
\nonu\\
&+& \h \( \psi \( \psi + 2 \d \) - \l_1 \( \l_1 + 2 \d\) -
\l_{M-1} \( \l_{M-1} + 2 \d\) \)  \mid P_{i} \rangle \otimes \mid P_{-j}
\rangle \nonu\\
&=& -{1\o{K+1}} \, \mid P_{i} \rangle \otimes \mid P_{-j} \rangle
\lab{adj1}
\er

One can easily check that
\be
\mid P_{j} \rangle \otimes \mid P_{-j} \rangle =
{1\o M} \mid S \rangle + \mid X_j \rangle
\ee
where
\br
\mid X_j \rangle &=& -{1\o M} \( \sum_{k=1}^{j-1} k \mid v_k \rangle
- \sum_{k=j}^{M-1} \( M-k\) \mid v_k \rangle \)
\er
and where we have denoted
$\mid v_k \rangle \equiv \( \mid P_{k} \rangle \otimes \mid P_{-k}
\rangle - \mid P_{k+1} \rangle \otimes \mid P_{-(k+1)} \rangle \)$.

Therefore
\br
&&\IC \,  \mid P_{j} \rangle \otimes \mid P_{-j} \rangle =
{1\o M}\( - {{M+K+1}\o{M(K+1)}} - \h \(  \l_1 \( \l_1 + 2 \d\) +
\l_{M-1} \( \l_{M-1} + 2 \d\) \) \) \mid S \rangle \nonu\\
&+&\! \( - {{M+K+1}\o{M(K+1)}} + \h \( \psi \( \psi + 2 \d \) -
\l_1 \( \l_1 + 2 \d\) - \l_{M-1} \( \l_{M-1} + 2 \d\) \) \)
\mid X_j \rangle \nonu\\
&=& -{1\o{K+1}} \mid P_{j} \rangle \otimes \mid P_{-j} \rangle -
\mid S \rangle
\lab{adj2}
\er
Consequently, from \rf{adj1} and \rf{adj2}
\be
\IC \, \mid P_{i} \rangle \otimes \mid P_{-j} \rangle =
-{1\o{K+1}} \mid P_{i} \rangle \otimes \mid P_{-j} \rangle -
\d_{ij} \sum_{k=1}^M \mid P_{k} \rangle  \otimes \mid P_{-k} \rangle
\lab{adj}
\ee

In addition one has
\be
\IC \, \mid P_{\pm i} \rangle \otimes \mid \a_a\cdot H^{(0)} \rangle =
\IC \, \mid \a_a\cdot H^{(0)} \rangle \otimes \mid \a_b\cdot H^{(0)} \rangle =
0
\lab{scalar}
\ee
with $a,b=M+1, M+2, \ldots, M+K$, $i=1,2,\ldots, M$.

{}From \rf{a20}, \rf{51a} and
$\Tr \( E_{\a}\, E_{-\a}\)=1$,
$\Tr \( \a_a \cdot H^{(0)}\, \a_b \cdot H^{(0)} \) = \a_a \cdot \a_b$
one gets that
\be
q_i = \vp \( P_{-i}\) \, , \qquad r_i = \vp \( P_{i}\) \, , \qquad
U_a = K_{ab}^{-1} \vp \( \a_b \cdot H^{(0)}\)
\ee
where $K_{ab}^{-1}$ is the inverse of $K_{ab} = \a_a \cdot \a_b$,  $a,b = M+1,
M+2, \ldots, M+K$.

Therefore from \rf{dirac}, \rf{diracterm}, \rf{symm1}, \rf{symm2}, \rf{adj} and
\rf{scalar} one gets the Dirac bracket for $sl (M+K+1)$,
which reproduces (after an appropriate Miura transformation)
the second bracket of {\sf cKP}$_{(K+1,M)}$ hierarchy:
\br
\pbr{r_i(x)}{q_j(y)} \eq \( \pa_x - U_{M+1}(x) -
\sum_{s=1}^{M} r_s(x) \pa^{-1}_x q_s(x) \) \d_{ij} \d (x-y)\nonu\\
&-& {1 \o K+1}  r_i(x) \pa^{-1}_x q_j(x)   \d (x-y) \lab{riqj} \\
\pbr{r_i(x)}{r_j(y)} \eq  {1 \o K+1}  r_i(x) \pa^{-1}_x r_j(x) \d (x-y)
+  r_j(x) \pa^{-1}_x r_i(x)  \d (x-y)      \lab{rirj} \\
\pbr{q_i(x)}{q_j(y)} \eq  {1 \o K+1}  q_i(x) \pa^{-1}_x q_j(x) \d (x-y)
+  q_j(x) \pa^{-1}_x q_i(x) \d (x-y)       \lab{qiqj} \\
\pbr{q_i(x)}{U_b(y)} \eq  - \h q_i(x) \d_{b M+1} \d (x-y)   \; ;\;
\pbr{r_i(x)}{U_b(y)} =   \h r_i(x) \d_{b M+1}    \d (x-y)
\phantom{aa}  \lab{qriub} \\
\pbr{U_a(x)}{U_b(y)} \eq   {1 \o 4}   K_{ab} \pa_x \d (x-y)      \lab{uaub}
\er

Note that for $K=0$ (and $U_a =0$) we recover from the above bracket
structure the second bracket of the NLS-$sl(M+1)$ model \ct{FK83,AGZ}.
This can also be checked directly by applying the same technique as above
to the model described in subsection 3.1.1.
Calculation shows that the equation \rf{casimirslmu1} in this case
is replaced by
\be
\IC =  {\tilde {\IC}}\,\,\, +\,\,\, {{M+1}\o {M}}\,\, \l_M\cdot
H^{(0)}\otimes \l_M\cdot H^{(0)}
\lab{casimirslmu2}
\ee
and equations \rf{symm1},\rf{symm2} and \rf{adj} hold with $K=0$.

For the purpose of illustration let us consider the example  of $sl (3)$
with $M=K=1$ corresponding to \cKP$_{2,1}$ with the Lax operator
as in \rf{lj}
\be
L_1 = (D - v_{2} ) \( D - v_{1} - r D^{-1} q \) =
D^2 + u + \P D^{-1} \Psi
\lab{ljsl3}
\ee
where $u = - v^2 - v^{\pr} - rq, \P= - r^{\pr} - rv, \Psi = q$ with
$v= v_{2} = - v_{1} $. The second bracket for $u, \P,\Psi$ obtained from
\rf{riqj}-\rf{uaub} indeed reproduces the standard \cKP$_{2,1}$ second bracket
(see e.g. \ct{cheng}).
One checks easily that the following equations:
\br
\pa_{t_2} q \eq  - q^{\pr \pr} + v^{\pr} q + r q^2 + v^2 q \nonu \\
\pa_{t_2} r \eq   r^{\pr \pr} + v^{\pr}r  + r^2 q + v^2 r \lab{flow2} \\
\pa_{t_2} v \eq  (rq)^{\pr} \nonu
\er
following from the Dirac bracket \rf{riqj}-\rf{uaub}  and the Hamiltonian
$H_1 = \int \P \Psi$ reproduce the correct flows for $u, \P,\Psi$:
\br
\partder{\Phi}{t_2} &=& {\pa^2 \Phi \o \pa x^2} + 2 u_0 \Phi
\nonu \\
\partder{\Psi}{t_2} &=& -{\pa^2 \Psi \o \pa x^2} -2 u_0 \Psi \lab{twored}\\
\partder{u_0}{t_2} &=& \pa_x (\Phi \Psi)
\nonu
\er
Equation \rf{twored} agrees with the second flow equation of the so-called
Yajima-Oikawa hierarchy \ct{konop,cheng}.
Note that in this calculation of the $t_2$ flow the consistency
of the Z-S problem required use of $b^{(2)} = \l^2 (\l_1 \cdot H)$ in \rf{B0}
according to the discussion of section 3.
Generally for $sl (3)$ we need to take
$b^{(2k)} = \l^{2k} (\l_1 \cdot H)$  and $b^{(2k+1)} = \l^{2k+1} E$.

\sect{Discussion and Outlook}
\label{sec:final}

In the formalism based on the pseudo-differential Lax operator,
the \cKP hierarchy is obtained by constraining the complete KP hierarchy
with the symmetry constraints expressed
in terms of the eigenfunctions $\P_i, \Psi_i$ from \rf{f-5}
and imposed on the isospectral flows.

In this paper we have obtained an alternative derivation of the \cKP
hierarchy as an integrable ${\widehat {sl}} (M+K+1)$-matrix hierarchy
generalizing the Drinfeld-Sokolov hierarchy.
The main ingredients of this construction were the semisimple graded,
non-regular element $E$ of $sl (M+K+1)$ and the potential $A$ belonging to the
grade zero subalgebra $\cgh_0$.
Both the Lax matrix operator as well as the underlying
recurrence operator were constructed in terms of these basic elements.
The matrix hierarchy exhibited a gauge symmetry related to $\kere$.
Due to presence of this gauge symmetry the relevant phase space turned out
to be the quotient space $\cgh_0 /\( \kere \cap \cgh_0\)$.
The structure of the flows of the hierarchy was shown to be related
to the center of $\kere$ \ct{wilson}.
The algebraic approach allowed us to write down in closed form a very
simple expression for the second Hamiltonian
structure with respect to which the flows are Hamiltonian.
This bracket structure was explicitly calculated as a Dirac bracket
emerging from reduction of $\cgh_0$ to the effective phase space of
$\cgh_0  / \(\kere \cap \cgh_0\)$.

One expects that several aspects of the \cKP formalism will gain substantially
by being treated  within the algebraic formalism proposed in this paper.
Work is in fact in progress regarding the following issues.
Possible extensions of the \cKP scalar Lax examples by going beyond the
algebraic construction based on the $sl(M)$  algebra by employing different
algebras.
Calculation of the tau-function, following the dressing transformation
method \ct{fhms} and Darboux-Backlund methods \ct{ANP2}
will help to further establish the connection between
the pseudo-differential and matrix techniques.
One also expects that the use of the matrix hierarchy
will be essential for describing additional symmetries of the \cKP models.

\lskip
{\bf Acknowledgements }
HA thanks Fapesp for financial support and IFT-Unesp for hospitality.
\lskip

\end{document}